# Gutenberg-Richter Law for Internetquakes

Sumiyoshi Abe[1] and Norikazu Suzuki[2]

[1]Institute of Physics, University of Tsukuba, Ibaraki 305-8571, Japan

[2]College of Science and Technology, Nihon University, Chiba 274-8501, Japan

The Internet is a complex system, which has intricate tangle, connection diversity, self-organization, and cluster and hierarchical structures. It is of current general interest to understand scale free properties of the Internet. Recently, several intriguing power-law distributions were reported for connectivity *(1)*, the number of links *(2)* and web pages *(3)* in the World-Wide Web. These scale free properties are essential in order that the network is resilient and robust to random errors, breakdowns, and attacks *(4)*.

To identify new properties of the Internet, we have performed a series of Ping experiments. Ping signals are emitted one after another from a local computer with a fixed time interval of 1s and make the round trip between the local computer and a destination host (i.e., the site accessed) through a number of routers. The resulting collection of round-trip times defines a time series, which shows the temporal behavior of the network. The Internet time series thus obtained is found to be highly nonstationary and to have two separate time scales: one is the user's long time scale (typically ~1hour) and the other is a short time scale (~100ms) associated with sudden drastic changes of the temporal pattern, referred to here as "Internetquakes". This nomenclature is chosen because of the fact that earthquakes exhibit identical temporal behavior: that is, external driving or energy injection may last for many years and the relaxation requires a few seconds or minutes.

Here, we present experimental evidence that the Gutenberg-Richter law *(5)* holds also for Internetquakes. This discovery confirms the scale free nature of the Internet.

The Gutenberg-Richter scaling law is an empirical law, which states that the logarithm of the cumulative frequency $N(>m)$ of earthquakes with magnitude larger than $m$ is proportional to the magnitude: $\log_{10} N(>m) = a - bm$, where $a$ and $b$ are positive constant. In seismology, magnitude is traditionally expressed in terms of the seismic moment $M$ as $m = \frac{1}{1.5} \log_{10} M - c$ with $c$ a constant close to 10, which leads to the moment-frequency scaling $N(>m) \sim M^{-b/1.5}$. Observations show that $b \sim 1$.

In the case of the Internet, "magnitude" $\mu$ of an Internetquake is the logarithm of round-trip time $\tau$: $\mu = \log_{10} \tau + \gamma$, where the constant $\gamma$ can simply be set equal to zero. In other words, the round-trip time corresponds to the seismic moment.

Following extensive tests, we have ascertained that there certainly exists a universal Gutenberg-Richter-type scaling behavior between the moment $\tau$ and the cumulative frequency $\nu(>\mu)$: $\nu(>\mu) \sim \tau^{-\beta}$.

In Fig. 1A, we present a small portion of a typical example of observed long time series data of round-trip time. The nonstationary behavior can clearly be observed. The larger the value of the round-trip time is, the more the network is congested. In Fig. 1B, we present the semi-log plot of magnitude and the cumulative frequency $\nu(>\mu)$. A clear scaling regime is recognized there. Through other extensive tests as well as the present one, the exponent $\beta$ is found so far to range between 1 and 6.

In conclusion, we have discovered the Gutenberg-Richter law for Internetquakes, thus demonstrating the scale free nature of the Internet. This suggests that, as complex systems, both earthquakes and the Internet share common features. In fact, according to our recent preliminary analysis, the Omori law, for the temporal pattern of aftershocks, is also valid for Internetquakes. An important point arising from this result is that the study of Internetquakes can shed new light on the study of earthquakes, with the obvious advantage that experiments on the former are much easier to perform than on the latter.

# Figure caption

**Fig. 1A**  A small portion of an observed long Internet time series. From the local host computer, ns2.phys.ge.cst.nihon-u.ac.jp (133.43.112.51), to the destination host, www.th.phys.titech.ac.jp (131.112.122.76), through 16 routers. The whole time series was measured continuously between 18:49 on 1 May, 2002 and 18:30 on 23 May, 2002. The total number of data values taken is 1648451.

**Fig. 1B**  The semi-log plot of magnitude and the cumulative frequency. An Internetquake is defined in such a way that its magnitude is smaller than the threshold value $\mu_{th} = \log_{10} \tau_{th}$. Here, $\tau_{th}$ is taken to be $\tau_{th} = 1 \times 10^3$ ms. The threshold allows one to distinguish the Internetquakes from isolated large fluctuations.

**A**

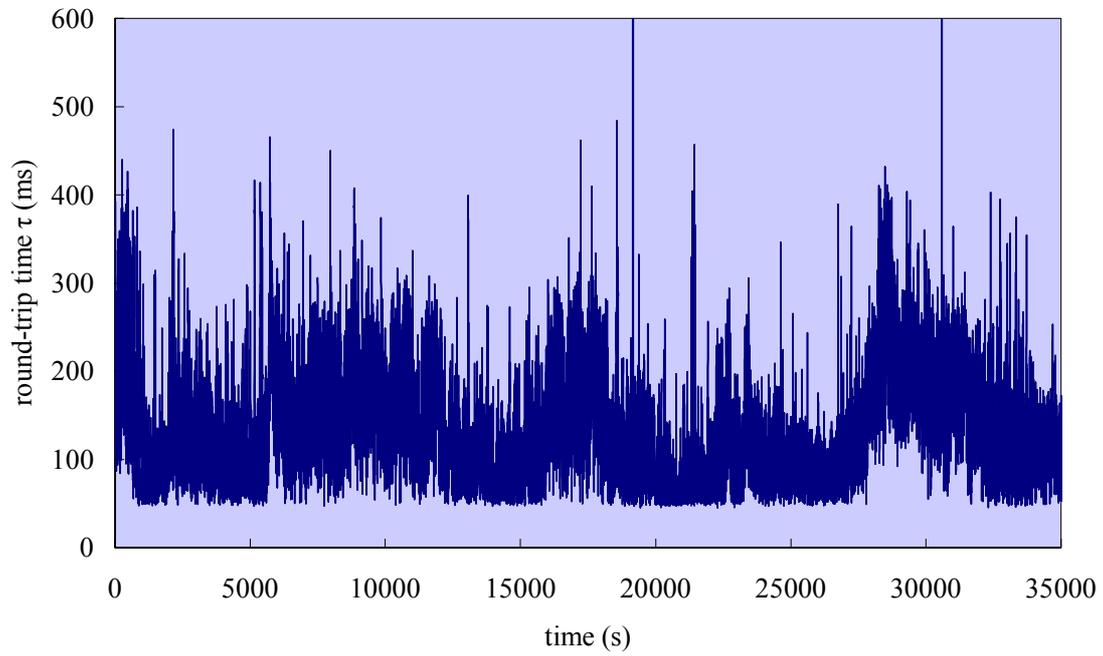

**B**

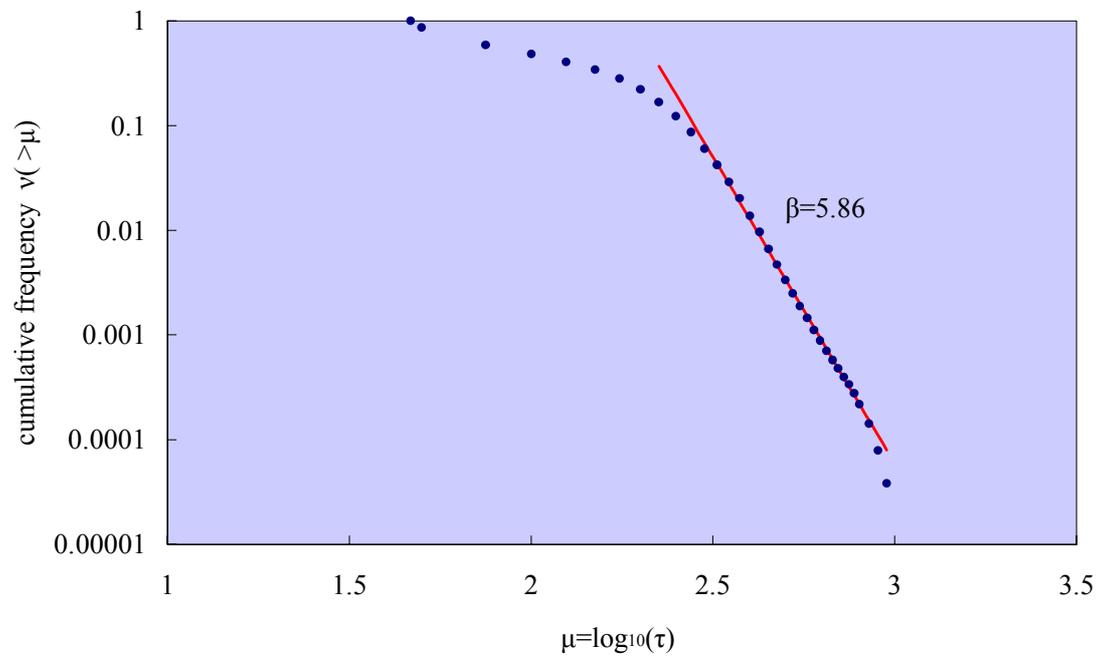

**Fig. 1**